\documentclass[conference]{IEEEtran}
\pdfoutput=1
\usepackage{placeins}
\usepackage{float}
\usepackage{graphicx}
\graphicspath{{./figs}}
\usepackage{cite}

\ifCLASSINFOpdf
\else
\fi
\usepackage[cmex10]{amsmath}
\newlength \figwidth
\if@twocolumn
\else
\fi
\IEEEoverridecommandlockouts
\begin{document}
%
\title{Adaptive Backoff Algorithm for IEEE 802.11 DCF under MPR Wireless Channels}

\author{\IEEEauthorblockN{Arun I B}
\IEEEauthorblockA{Department of Electrical Engineering\\
Indian Institute of Technology Madras\\
Chennai - 600036, India\\
Email:arunib@smail.iitm.ac.in}
\and
\IEEEauthorblockN{T.G. Venkatesh}
\IEEEauthorblockA{Department of Electrical Engineering\\
Indian Institute of Technology Madras\\
Chennai - 600036, India\\
Email:tgvenky@ee.iitm.ac.in}
\thanks{In proc. of ICCCN 2013, Nassau, Bahamas. July 30 - August 2, 2013.

\copyright 2013 IEEE. Personal use of this material is permitted. Permission from IEEE must be obtained for all other uses, in any current or future media, including reprinting/republishing this material for advertising or promotional purposes, creating new collective works, for resale or redistribution to servers or lists, or reuse of any copyrighted component of this work in other works}
}


%


\maketitle

\begin{abstract}
As a result of the recent advances in physical (PHY) layer communication techniques, it is possible to receive multiple packets at the receiver concurrently. This capability of a receiver to decode multiple simultaneous transmissions is known as multi-packet reception (MPR). In this paper, we propose a simple Medium Access Control (MAC) protocol for an MPR wireless channel, where we modify the backoff procedure as a function of number of ongoing transmissions in the channel. Our protocol is backward compatible with the IEEE 802.11 DCF protocol. The performance analysis of the proposed protocol is carried out using extensive simulations and it is compared with some of the existing MPR MAC protocols. The proposed mechanism improves the throughput and delay performance of the IEEE 802.11 DCF.

\end{abstract}

Keywords: WirelessLANs, 802.11 DCF, Multipacket Reception, Medium Access Sublayer, Simulation

%
\IEEEpeerreviewmaketitle

\section{Introduction}
\subsection{Background}
Traditional collision model for describing the wireless channel has been the norm, up until very recent times.
In a collision channel model, a packet can be received successfully at the receiver only if there is exactly one transmitter attempting to access the channel at a given slot. In other words, the channel could only be in three possible states, viz idle, busy or collision. However, due to the recent advances in physical (PHY) layer technologies, it is possible to receive or decode a transmission even in the presence of other transmissions. These effects can be classified as capture, in which case one out of the several simultaneous transmissions will be decoded or as multipacket reception (MPR) in which more than one transmission could be simultaneously decoded at the receiver. The detailed survey of the PHY layer technologies that makes MPR possible can be found in \cite{Lu2012}.
Traditional Medium Access Control (MAC)  protocols which were designed, keeping in mind the collision channel model, does not make use of the additional freedoms of the multipacket reception channel. Those traditional MAC protocols underestimate the channel capacity leading to inefficient use of the wireless channel. A common characteristic of MPR MAC protocols is that nodes are allowed to transmit even when the channel is sensed to be busy.
\subsection{Literature}
    Earliest reference to a multipacket reception channel dates back to the late 1980s. In their 1988 paper \cite{ghez1988}, Ghez \textit{et al.} describe an MPR channel model and study the stability properties of slotted aloha systems under such channels. 
Since then, several MAC protocols for MPR channel have been proposed in the literature. In 2001, Zhao, Tong \textit{et al.} \cite{qzhao2003} proposed a Multi-Queue Service Room (MQSR) protocol for the MPR channel. Later on the same authors\cite{qzhao2004} went on to propose simpler suboptimal dynamic queue protocol  as an improvement over the high computational complexity of MQSR protocol. Both these protocols require a centralized controller for coordinating node transmissions.
Chan and Berger \cite{dschan2004} proposed a cross-layer solution for MPR known as cross layer CSMA (XL-CSMA). It is a decentralized protocol in which station makes the decision to transmit based on information obtained from carrier sensing. It is assumed that stations can estimate the fraction of channel capacity used, based on PHY layer parameters.
A reservation based protocol called Multi Reservation Multiple Access (MRMA) was proposed by Hui Chen \textit{et al.}\cite{Chen2005a} in 2005. The authors propose a centralized reservation scheme for channel access which provides QoS (Quality of Service) for multimedia traffic.

The first MPR MAC protocol, which was based on the 802.11 Distributed Coordination Function (DCF) was proposed by Zheng and Angela\cite{pxzheng2006} in 2006. This protocol modifies the packet structure of the CTS and ACK of 802.11 DCF to accommodate acknowledging of multiple stations. 
Angela Zhang \cite{Junb} later proposed a multi-round contention protocol in which several rounds of contention take place before the data transmission in case the number of winners are small. Another protocol based on DCF was proposed by Barghi \textit{et al.}\cite{Barghi2011}. The authors propose a Multiple Input Multiple Output (MIMO) based cross layer design in which some changes are made to the RTS-CTS signaling. Mahmood \textit{et al.} \cite{Mahmood2010} proposed a modification of DCF in which contention window size is controlled according to network loads to obtain throughput gains. 
Recently Babich \textit{et al.} have proposed \cite{fbabich2010} a generalization of 802.11 DCF to the MPR channels. The authors propose a threshold based technique in which the backoff counter is frozen only when the number of ongoing transmissions in the channel is greater than a threshold. This protocol henceforth will be referred to as the threshold based protocol.

Although there has been several attempts in literature to design MPR aware MAC protocols, a robust backward compatible protocol for MPR channel is yet to emerge. The upcoming Wireless LAN standard 802.11ac, which supports optional MU-MIMO \cite{6359961}, an MPR enabling technology, uses Enhanced Distributed Channel Access (EDCA) for medium access. EDCAF (EDCA Function) is an extension of DCF to support priority traffic. However, EDCA protocol, not being designed as an MPR aware protocol, may not fully exploit the MPR capabilities of the channel. 
 
In this paper, we propose a simple adaptive backoff scheme for the 802.11 protocol to utilize the MPR capability more effectively. The proposed scheme operates entirely within the MAC layer and does not suffer from the drawbacks \cite{Kawadia2005} of cross layer designs. Our protocol operates in a fully distributed fashion and do not require any form of centralized coordination.  
  
  The remainder of the paper is organized as follows. Section \ref{sec:model} describes the MPR channel and network models. In Section \ref{sec:protocoldesign}, design of the proposed scheme is presented. In Section \ref{sec:perfeval}, performance evaluation of the protocol is presented. Section \ref{sec:conclusion} concludes the paper.
\section{System Model} \label{sec:model}
\subsection{Channel Models}

The MPR channel models which are widely used in literature are described below.

\subsubsection{Generalized MPR Channel} \label{subsubsec:genmprchannel}
In the generalized MPR channel due to Ghez \textit{et al.} \cite{ghez1988}, a node will be able to receive  $m$ out of $n$ transmissions with certain nonzero probability. More over the probability of successful reception depends only on $m$ and $n$.

If $i$ packets are transmitted in a slot, we define
\[
\epsilon_{ij} \equiv \Pr( j \mbox{ packets are received} | i \mbox{ are transmitted})
\]

Generalized MPR channel can be characterized by the probabilities $\epsilon_{ij}$ for all values of $i$ and $j$.
These $\epsilon_{ij}$ values can be summarized in  matrix form called \textbf{reception matrix} of the channel.

\[
E\equiv {
\left[\begin{array}{cccccc}
\epsilon_{10} & \epsilon_{11}\\
\epsilon_{20} & \epsilon_{21} & \epsilon_{22} & 0\\
. & . & \\
\epsilon_{n0} & \epsilon_{n1} & \epsilon_{n2}  \\
.&.&&&.&\\
.&.&&&&.\\
\end{array}\right]}
\]

\subsubsection{$k$ - MPR}
In a $k$ - MPR channel, a node will be able to receive all packets without loss whenever the number of packets transmitted is not greater than $k$.
In the event that number of transmissions goes above $k$, collision occurs and nodes will not be able to receive any of the packets. If $\zeta$ denotes the number of concurrent transmissions in a collision domain,
\[\Pr(\mbox{Success}) = \left\{
\begin{array}{l l}
  1 & \quad \mbox{if $\zeta \le k$}\\
  0 & \quad \mbox{if $\zeta >  k$}\\
\end{array} \right. \]
As can be noted, in a $k$-MPR channel, either all transmissions are successful or none of them are successful. Such a case can occur when successful reception directly depends on the interference level at the receiver (SINR). The $k$-MPR channel as well as the conventional collision channels are special cases of generalized MPR channel. The choice of a particular channel model for a specific scenario depends on the PHY layer technologies employed. 

\subsection{Network Model}
We model the network as an ad-hoc wireless network in which receivers operates in a completely distributed manner. Every node in the network is equipped with receivers capable of receiving up to $k$ transmissions concurrently without error. All nodes in the network are half-duplex; i.e., it is not possible for a node to transmit and receive simultaneously. A crucial requirement for our protocol to work is that of enhanced carrier sensing. That is, non-transmitting nodes have the capability to sense (or estimate accurately) the number of ongoing transmissions. This enhanced carrier sensing capacity is known as multi-dimensional \cite{Lin2011} carrier sensing or MIMO \cite{Coviello2009} carrier sensing. 

\section{Protocol Design} \label{sec:protocoldesign}

\subsection{Motivation}
PHY layer technologies which enable MPR, include Code Division Multiple Access (CDMA), Orthogonal Frequency Division Multiple Access (OFDMA), Multiuser MIMO (MU-MIMO) etc. A MAC protocol for multipacket reception can be built around the underlying physical layer technology, in which case it is said to be a \emph{”Cross Layer”} protocol. An alternative is to abstract out the PHY layer and confine the protocol to the MAC layer. We focus on the latter approach, as the protocol design will be simpler and can be used with all MPR capable devices, irrespective of the underlying physical layer implementation. 

The \textit{de facto} standard for WLAN medium access is the 802.11 DCF. Hence any MPR aware MAC protocol for the sake of backward compatibility should not differ widely from the DCF.
The basic DCF protocol, which is a CSMA/CA (Carrier Sense Multiple Access/Collision Avoidance) scheme with binary exponential backoff, is as described below \cite{6178212}.

\subsection{Review of 802.11 DCF and its adaptations for MPR}
In 802.11 DCF, a station which has a packet to transmit senses the medium to determine if its idle. Here, the wireless channel is slotted with slot length, say $\sigma$. If the channel is sensed to be idle without interruption for a duration equal to the DIFS, the station proceeds with transmission.
On the other hand, if the medium is sensed to be busy, the station waits till the end of the current transmission. Then it generates a random backoff value drawn uniformly from the interval $0$ to $CW_{min}$ (minimum contention window size) and continue to sense the channel. The backoff counter maintained by the node is decremented at the end of each idle slot. The countdown process is frozen whenever the channel becomes busy.
When the backoff counter reaches $0$, the node attempts transmission. If the transmission is successful (receipt of ACK frame), the next packet is processed. On the other hand if the packet transmission is unsuccessful, random backoff counter value is again generated (but with double the mean) and the countdown process begins. The process is continued until the transmission is successful or until maximum number of retries is reached upon which the packet is dropped.

In the conventional 802.11 DCF, whenever there is at least one ongoing transmission, the backoff counter is frozen to avoid collisions. However, an MPR system can support multiple transmissions. 
Hence the usage of conventional 802.11 DCF for MPR systems leads to underutilization of the channel capacity. Accordingly, some variations of the basic protocol has been proposed \cite{pxzheng2006,fbabich2010} which are better suited for the MPR scenario. The variations are in the backoff process, specifically in the decrementing of the backoff counter.
In a variation known as the threshold based protocol due to Babich \cite{fbabich2010}, the backoff counter is frozen whenever the number of ongoing transmissions is greater than a threshold ($L_t$). In this protocol, stations are allowed to decrement the backoff counter by unity, whenever a slot elapses in which the number of transmissions is less than or equal to $L_t$. The threshold $L_t$ is usually set to be less than the MPR capability of the channel. 
\subsection{Adaptive backoff algorithm}
 In our proposed protocol, the backoff counter is frozen only when the number of ongoing transmissions is greater than or equal to a threshold. The value of the threshold can be fixed to be equal to or less than the MPR capability of the node. Further, the backoff counter will be decremented by the number of additional possible transmissions. If the MPR capability is $K$ and there are $i$ ongoing transmissions, the backoff counter shall be decremented by $K-i$. This technique makes better use of the feedback regarding the channel utilization. If the number of transmissions are less, the counter gets decremented faster and the transmissions take place sooner leading to more aggressive channel access.
Our protocol design is inspired by the fact that the performance of DCF protocol can be improved by making use of the knowledge on the number of ongoing transmissions. Typically, the number of ongoing transmissions is a rough indicator of the prevailing traffic conditions. Therefore, if the backoff process were made a function of the number of ongoing transmissions, the throughput can be improved and delay decreased. 
It has been shown (for non MPR channels) by Bianchi \textit{et al.} \cite{Bianchi1996a} that 802.11 performance can be improved by employing an adaptive contention window based scheme. Hence similar performance improvements can be expected from MPR channels too by making use of an adaptive scheme and performance close to the channel capacity can be achieved.

If we denote $d(i)$ as the amount by which the backoff counter is decremented when a slot time elapses in which $i$ transmissions are going on, we get for conventional DCF,
\[d(i) = \left\{
\begin{array}{l l}
  1 & \quad \mbox{$i=0$}\\
  0 & \quad \mbox{otherwise}\\
\end{array} \right. \]
For threshold based protocol, 
\[d(i) = \left\{
\begin{array}{l l}
  1 & \quad \mbox{$i\le L_t$}\\
  0 & \quad \mbox{otherwise}\\
\end{array} \right. \]
For our adaptive MPR protocol,
\[d(i) = \left\{
\begin{array}{l l}
  K-i & \quad \mbox{$i \le K_t$}\\
  0 & \quad \mbox{otherwise}\\
\end{array} \right. \]
where $K_t$ ($<K$) is the threshold and $K$ is the MPR limit.

\subsubsection*{Estimates of $K$ for Generalized MPR Channel}
The above description of the protocol is based on a $k$-MPR channel model. For the case of generalized MPR channel, the same protocol can be used by appropriately defining an MPR capability $K$.
Intuitively $K$ should be the number of ongoing transmissions which achieves the maximum number of successful reception. When $i$ transmissions are attempted, the expected number of packets successfully received in a generalized MPR channel with the reception matrix $E$ given in \ref{subsubsec:genmprchannel} is given by $\sum_{j} j\epsilon_{ij}$. Therefore, the equivalent MPR capability for the channel can be computed as,
\begin{equation} \label{eq:genmprk}
K_{{equiv}} = \arg\max_i \sum_{j} j\epsilon_{ij}  
\end{equation}
If the RHS of \eqref{eq:genmprk} is not unique, minimum value can be used to save transmit power \cite{qzhao2004} or a middle value may be chosen to improve throughput.
\subsection{Protocol Details}
The definitions of SIFS (Short Inter Frame Space), DIFS (DCF Inter Frame Space), EIFS (Extended Inter Frame Space) for our protocol, remains the same as defined in the IEEE 802.11 standard \cite{6178212}. The difference comes under the conditions in which the MAC layer sees the channel as \textquotedblleft Idle" or \textquotedblleft Busy". In the proposed protocol, an \textquotedblleft Idle Slot" is one in which number of ongoing transmissions is less than or equal to the threshold $K_t$. A slot is \textquotedblleft Busy" only if there are at least $K_t+1$ ongoing transmissions in the channel. Since the decrements of the counter values in our protocol can be by a number greater than one, the counter may reach negative values without ever reaching zero. Therefore the nodes should transmit as soon as the counter reaches a negative value or zero. Whenever a node attempts to access the channel, it waits for a duration of DIFS in which no more than $K_t$ transmissions take place. In contrast, the conventional DCF and the threshold based protocol require the channel to be completely idle with no transmissions. A consequence of not freezing the counter during an ongoing transmission is that, unlike in the case of conventional DCF, the transmissions from different nodes will not be frame synchronous. Further, some of the ongoing transmissions may encounter collisions due to the following reason.

Even though there are some ongoing transmissions, two or more nodes can reach a non-positive counter value at the same slot  and may begin their transmissions. If the number of ongoing transmissions and the newly initiated transmissions exceeds the MPR limit, a  collision will occur not only for the newly initiated transmission but also for the existing transmissions.
  As a result, a transmission can be concluded to be successful only if it does not encounter collision till its completion.

\subsection{Backwards Compatibility} \label{subsec:compatibility}
The proposed protocol shares the basic structure of binary exponential backoff with conventional DCF, which ensures its backward compatibility. The devices operating with proposed protocol can co-exist with legacy 802.11 devices. 
The proposed backoff mechanism can be used to enhance the performance of EDCA protocol of 802.11ac. Further, we propose the use of different thresholds as a means to offer differential services. High priority packets can set a high value for threshold while lower priority packets use smaller values as threshold. 
In fact, the proposed backoff enhancement can be applied to a wide class of CSMA/CA protocols provided the enhanced carrier sensing assumption is valid. 
\subsection{Performance Considerations} \label{subsec:performance}
The performance of our protocol depends on the channel and traffic characteristics. It also depends on the values of the protocol parameters being used. When the nodes are all saturated, the effect of decrementing the backoff counter by a value greater than one, is equivalent to reducing the size of contention window. Hence in that case, the throughput performance of the proposed protocol will be similar to the threshold based protocol. However, this should not be a concern given that most of the time real networks operate under unsaturated traffic \cite{malone2007}. In the unsaturated case, under time varying traffic, our protocol should outperform the threshold based protocol. The proposed protocol, being an adaptive protocol, quickly adjusts itself to the variations in the offered traffic. Similarly, the proposed protocol can adjust itself abruptly to changes in the number of contending stations. On the other hand, when the traffic properties remain the same for a long time, the performance falls back to non-optimal values.

\section{Performance Evaluation} \label{sec:perfeval}
\subsection{Simulation Setup}
The network model adopted corresponds to an $N$ user uplink to a common access point. $N$ Nodes are communicating with a central node having MPR capability of $K$, through a common wireless channel. This situation is equivalent to an ad-hoc network of $N$ stations, each having MPR capability of $K$, where data exchange takes place between two arbitrary stations.
All nodes are assumed to have an infinite buffer. The packet arrival processes at each node is a Poisson process independent of packet arrivals at other nodes. The packet arrival rates to all nodes are equal. Further, all packets are of fixed size. We assume $k$-MPR channel model in our simulations, although the proposed protocol can be used under generalized MPR channels. Lastly, we assume ideal channel conditions - i.e. transmission errors occur only as a result of collisions. The simulations were carried out for basic access only (No RTS/CTS).

In our simulation, we have used the network parameters given in Table \ref{tab_simparams}, mostly taken from 802.11 standard \cite{6178212} for FH-PHY.
\begin{table}[h!]
\renewcommand{\arraystretch}{1}
\caption{Simulation Parameters \label{tab_simparams}}
\centering
\begin{tabular}{ l l }
\hline
    \hline
  Parameter & Value \\
  \hline
  Packet payload & 8184  bits \\
  MAC Header & 272 bits \\
  PHY Header & 128 bits \\
  Channel Bit Rate & 1Mbps \\
  Slot time ($\sigma$) & 50 $\mu s$ \\
  DIFS & 128  $\mu s$ \\
  Max backoff stage ($m$) & 5 \\
  Retry limit  & 4 \\
  \hline
    \hline
\end{tabular}
\end{table}
The simulations are done using SimPy \cite{SimPy} discrete event simulator, using Python. Since our goal was to study the MAC layer performance of different protocols, it was important to decouple the MAC from PHY layer implementation details. Therefore a simulator with an ideal physical layer was implemented using SimPy discrete event simulator. 

\subsection{Simulation Results} \label{subsec:results}
We have simulated the performance of IEEE 802.11 protocol as well as its backoff variants under differing MPR capabilities. The throughput and the delay of the protocols are obtained. 

\subsubsection{Throughput}
We use the notion of capacity of the channel as the maximum data rate supported  by one stream (i.e. one orthogonal code in the case of CDMA or one spatial stream for MU-MIMO). We define normalized throughput as the fraction of channel capacity used for actual data transmission.  This implies that a system with MPR limit $K$ can have throughput values ranging from $0$ to $K$. It may be noted that the throughput described above is MAC layer throughput obtained under ideal PHY layer. The actual throughput obtained at higher layers can be less.

\begin{figure}
\centering
\includegraphics[width=\figwidth]{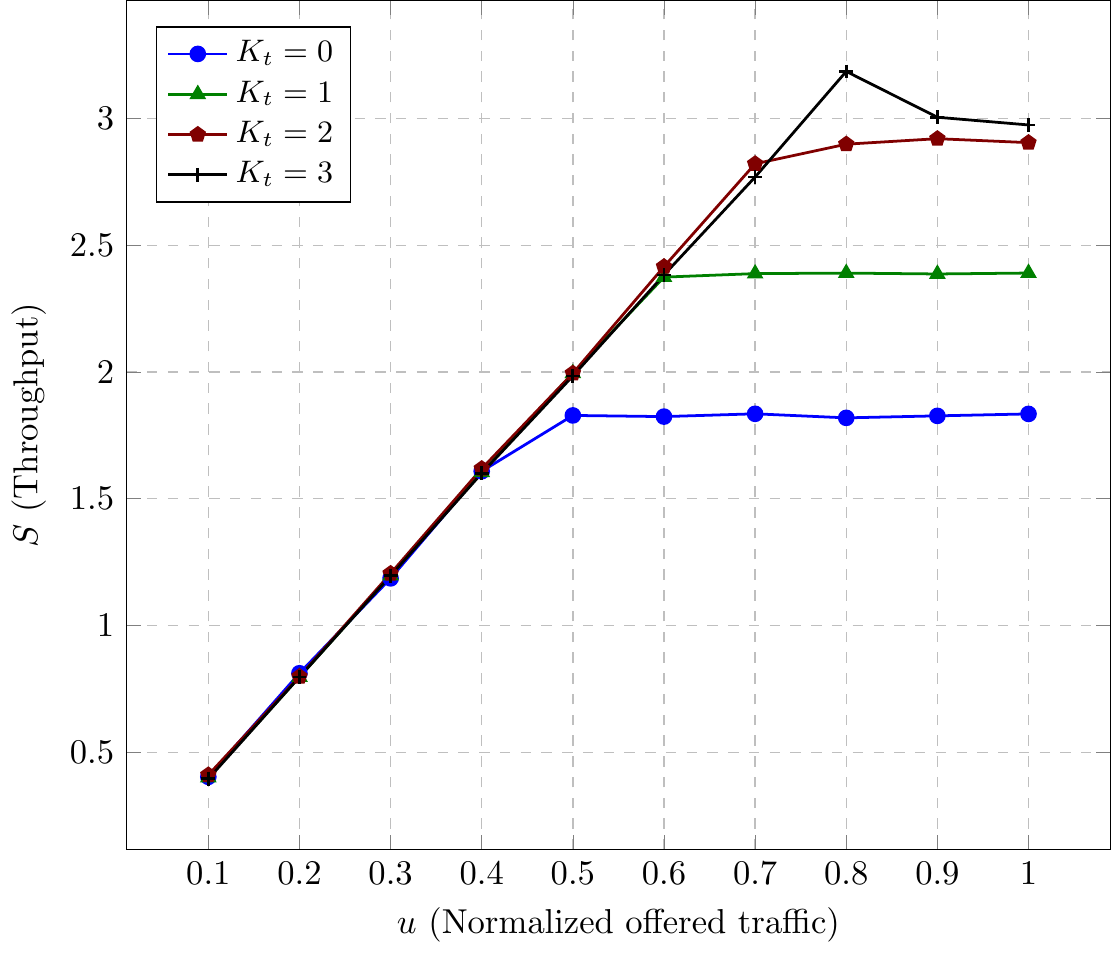}
\caption{The throughput of proposed protocol against normalized offered traffic (\emph{Params}: MPR limit $K=4$, Number of stations N = 30, Minimum contention window size $CW_{min}=128$)} 
\label{fig:uallproposedutpn30k4}
\end{figure}

\begin{figure}
\centering
\includegraphics[width=\figwidth]{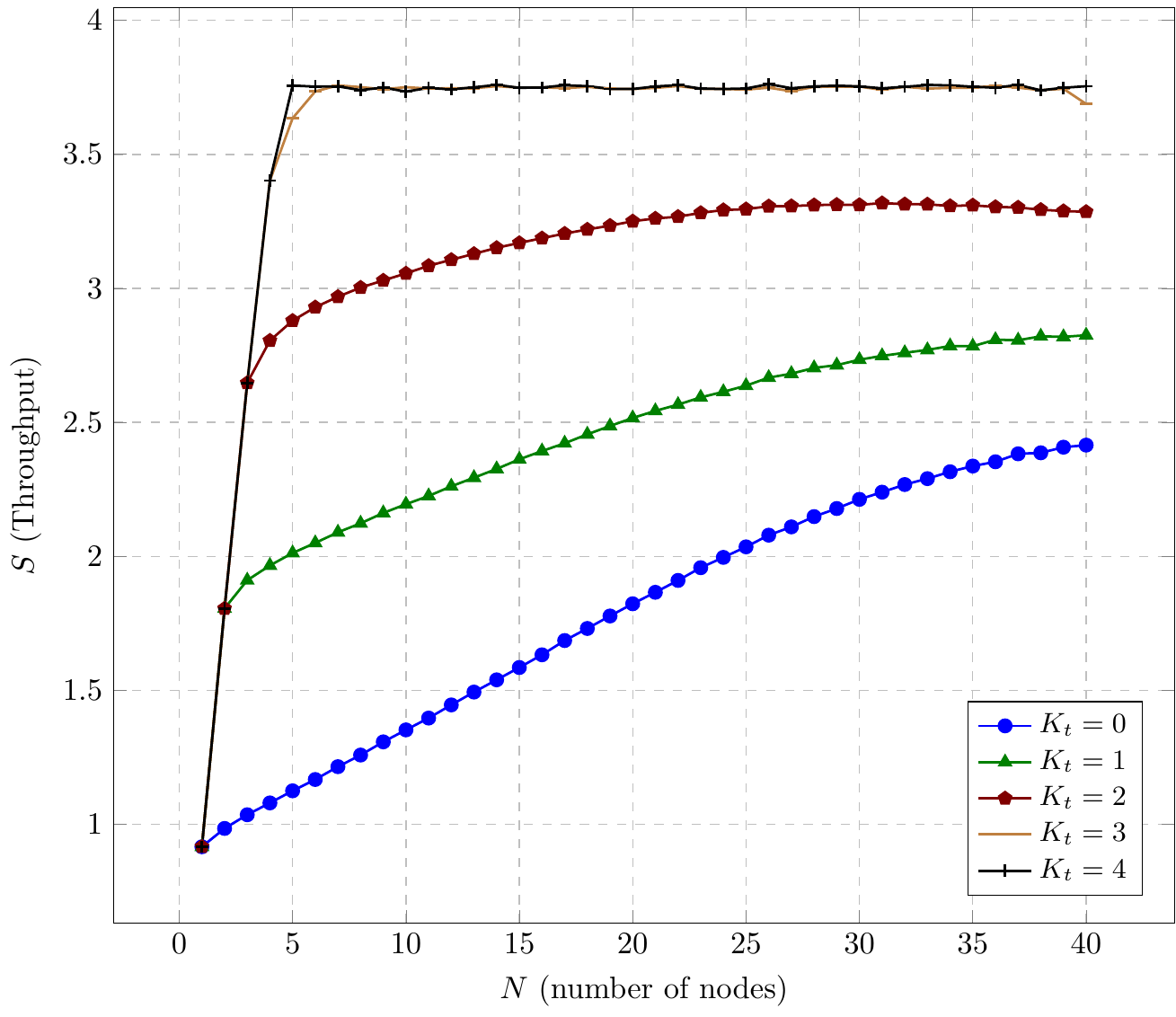}
\caption{The total normalized MAC throughput (unsaturated) of proposed protocol against number of nodes for different thresholds (\emph{Params}: MPR limit $K=5$, Normalized offered traffic = 0.75, Min. contention window size $CW_{min}=128$)}
\label{fig:utpproposedt} 
\end{figure}
\begin{figure}
\centering
\includegraphics[width=\figwidth]{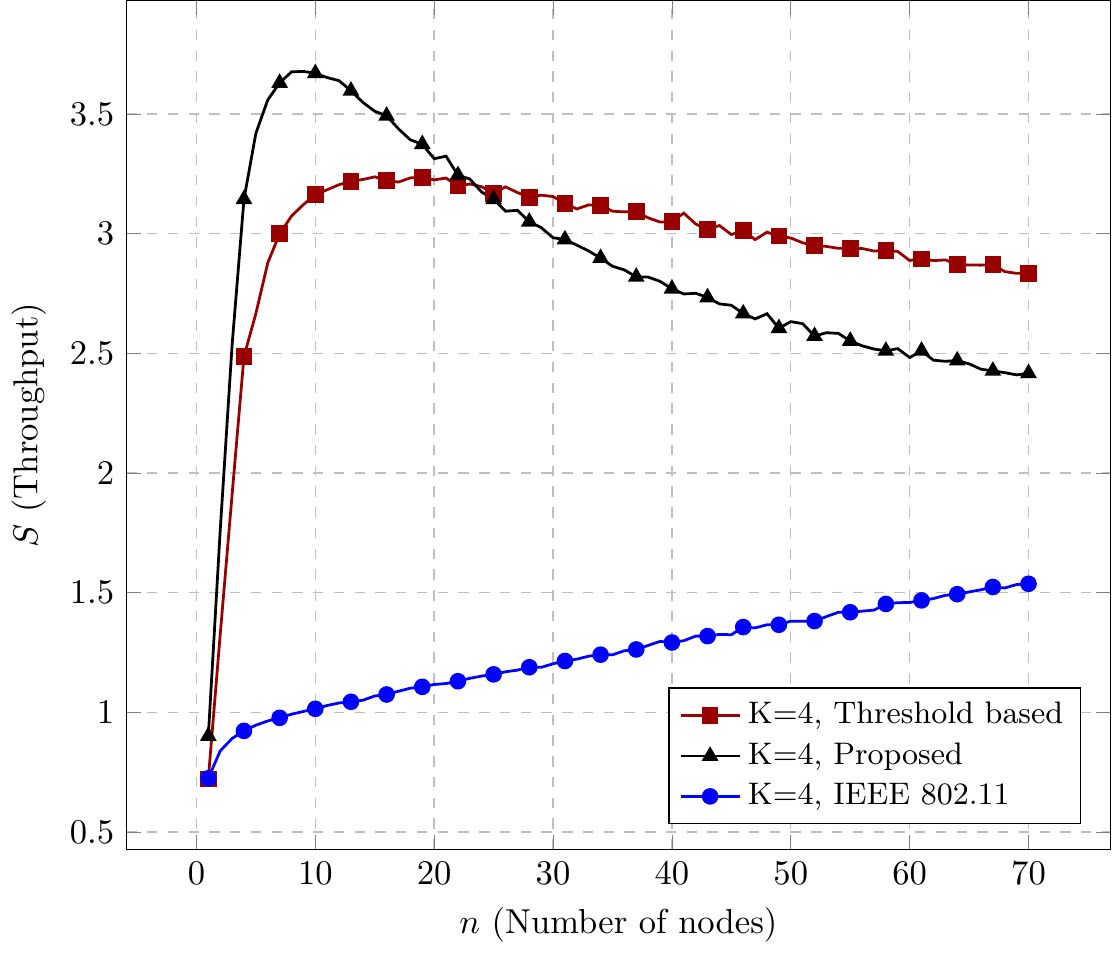}
\caption{The normalized saturation throughput of different protocols against number of nodes (\emph{Params}: MPR limit $K=4$, Normalized offered traffic = 1, Minimum contention window size $CW_{min}=128$)} 
\label{fig:unsattp_protocols}
\end{figure}
In \figurename \ref{fig:uallproposedutpn30k4} the aggregate (of all nodes) MAC throughput  of the proposed protocol is plotted against normalized offered traffic (total offered traffic measured in terms of channel capacity), for different values of thresholds. When the offered traffic is small, the throughput increases linearly with offered traffic. As the threshold is increased from $0$ to  $K-1$, the throughput increases as expected.  When the threshold is low, the throughput is limited by the prohibitive MAC policy whereas at higher thresholds increase in collisions limits the throughput.

The unsaturated throughput of the proposed protocol is plotted against number of nodes for different values of threshold $K_t$ in \figurename \ref{fig:utpproposedt}. For low values of threshold, as the threshold is increased, the throughput also increases. However, when the threshold set is near the MPR limit, the probability of a new transmission interfering with an ongoing transmission is high and hence further increase in threshold does not result in throughput increase.

In contrast to the 802.11 DCF without MPR, multiple simultaneous transmissions are possible when MPR capability is available. For example, if two or more counters reach value zero simultaneously, all of them can result in successful transmissions as long as the number of transmissions is bounded by the MPR limit. Therefore, throughput of more than one, is achievable in the case of 802.11 DCF with MPR.

In \figurename \ref{fig:unsattp_protocols}, the normalized throughput is plotted against number of nodes. In conventional DCF, a node is allowed to transmit only when there are no other ongoing transmissions on the channel. This restriction leads to a gross underutilization of channel capacity as is evident from \figurename \ref{fig:unsattp_protocols}. When the number of stations are small, the throughput performance of the proposed protocol is better than that of the threshold based protocol. As the number of nodes increases, the variation in the aggregate attempt rate of the nodes will be small. Thus, as described in section \ref{subsec:performance}, the performance of an adaptive protocol will be same as that of a non adaptive protocol in such saturated case. Here the throughput performance of proposed protocol is slightly worse for saturated case due to the effects described in detail later.

\begin{figure}
\centering
\includegraphics[width=\figwidth]{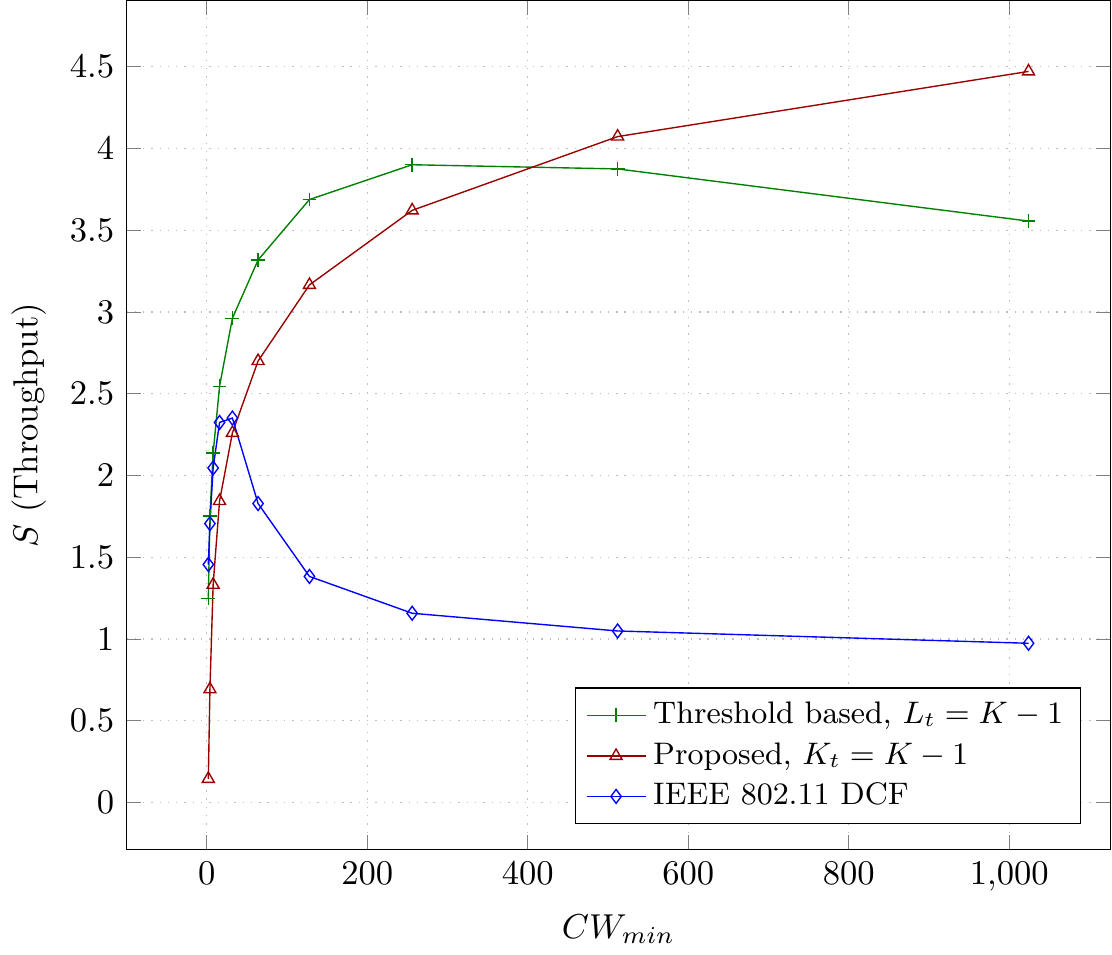}
\caption{The saturated MAC throughput of different protocols against minimum contention window size $CW_{min}$ (\emph{Params}: Number of stations $N=50$, MPR limit $K=5$,  normalized offered traffic $u=1$)} 
\label{fig:unsattp_cwmin100_n30}
\end{figure}

The variation of saturated throughput of the protocols with minimum contention window size ($CW_{min}$) is plotted in \figurename \ref{fig:unsattp_cwmin100_n30}. Due to the aggressive nature of the proposed protocol, its throughput is lower than that of threshold based protocol for small values of contention window sizes. As the $CW_{min}$ values are increased beyond a point ($CW_{min}\sim$ 400), the proposed protocol performs better than the threshold based protocol. It also points to the fact that $CW_{min}$ needs to be set to higher values to achieve performance gains from the proposed protocol. Further, the maximum throughput corresponding to optimum $CW_{min}$ value of for the proposed protocol is better than that of threshold based protocol. 

\subsubsection*{Optimal threshold}
The performance of our protocol is dependent on the value of the threshold set. This is illustrated by the plots in \figurename \ref{fig:uallproposedutpn30k4}, \figurename \ref{fig:utpproposedt}, and \figurename \ref{fig:uallproposedudelayn30k4}. We observe that as the threshold is increased towards MPR limit $K$, delay of the proposed protocol reduces and throughput increases. However, the performance of $K_t=2$ and $K_t=3$ are close, with $K_t=3$ having slightly better performance. Hence we set the threshold $K_t$ as $K-1$ for optimal performance in all our simulations wherever $K_t$ is fixed. In general the optimal threshold depends on the offered traffic, the number of contending stations, and protocol parameters.

\subsubsection*{Performance under saturated traffic conditions} \label{subsubsec:persat}
In 802.11 DCF and its variants (threshold based and proposed), the size of the contention window is reset to $CW_{min}$ after every successful transmission. In general, the optimal size of the contention window, for which the probability of successful transmission is maximum, lies between $CW_{min}$ and $CW_{max}$. This optimal value is higher for our proposed protocol (due to faster decrementing of the counter values) as compared to the threshold based protocol. As a result, the average number of collisions experienced by a packet before getting successfully transmitted is more for the proposed protocol than for the threshold based protocol in the saturation region. This is evident from \figurename \ref{fig:uall_combinedueff} wherein the (inverse of) number of transmission attempts per successful packet delivery is plotted against normalized offered traffic.
Further, due to the finite value of retry limit, more packets are likely to be dropped by the proposed protocol than by the threshold based protocol for smaller values of $CW_{max}$. Since dropped packets do not contribute to throughput, the throughput of the proposed protocol is observed to be less than that of the threshold based protocol, for the values of parameters fixed for the simulation shown in \figurename \ref{fig:unsattp_protocols}. Thus, we conclude that the throughput performance of the proposed protocol is better than that of the threshold based protocol only in the unsaturated conditions which depicts the case of most real networks.

\subsubsection{MAC Delay}
The medium access delay for a packet includes the time spent in collisions as well as the time spent in backoff process. The dropped packets pose problems to delay calculation. Clearly, the use of theoretical value of infinity, to account for the delay of a dropped packet is not beneficial. An alternative is to exclude all dropped packets from delay calculation. However, this latter approach does not give a true picture of the MAC layer delay performance. Therefore we calculate MAC delay as the time elapsed between the moment a packet is put to service (Head of Queue) and its successful transmission or drop (upon reaching the retransmission limit).

\begin{figure}
\centering
\includegraphics[width=\figwidth]{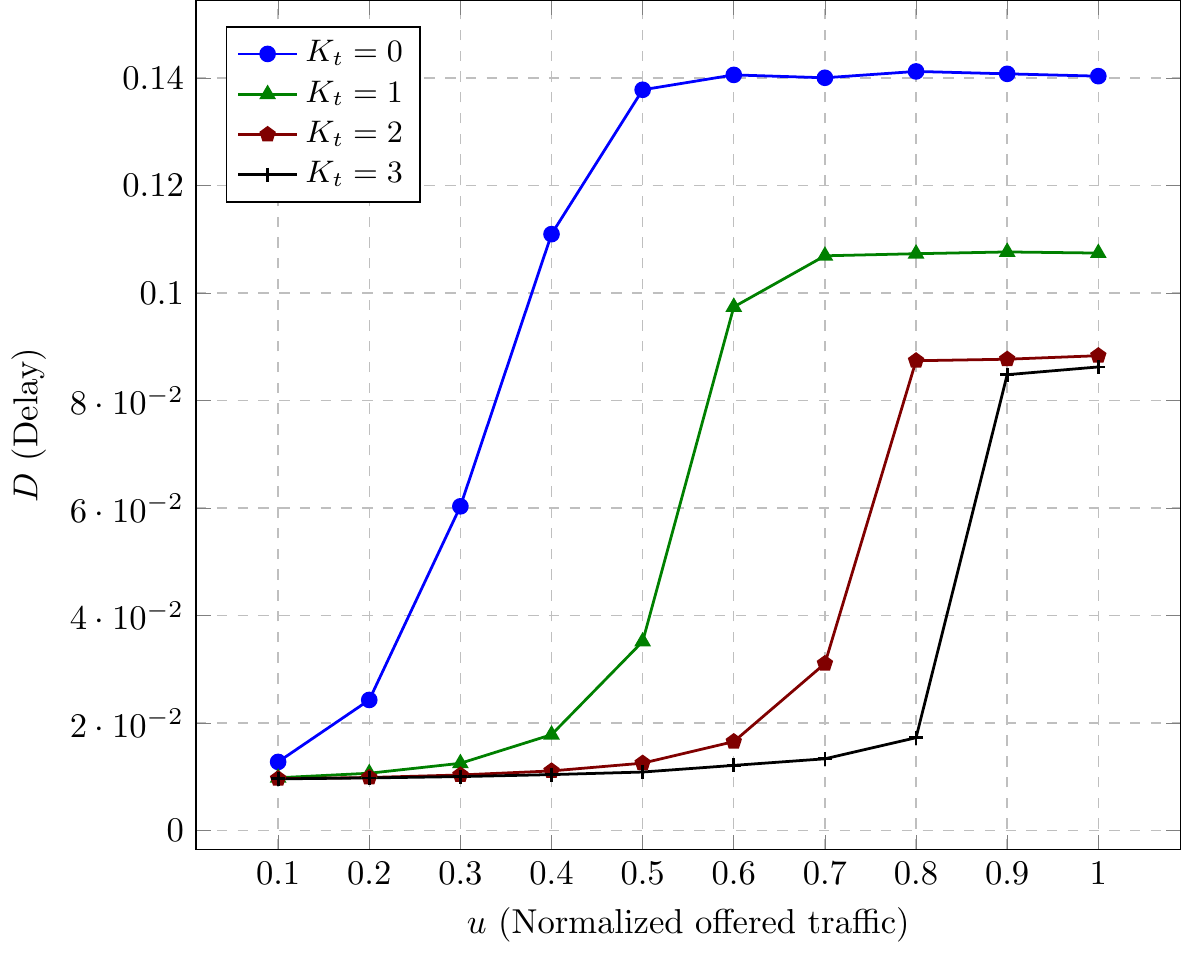}
\caption{The average MAC delay of proposed protocol against normalized offered traffic for different values of threshold (\emph{Params}: MPR limit $K=4$, Number of stations N = 30, Min. contention window size $CW_{min}=128$)} 
\label{fig:uallproposedudelayn30k4}
\end{figure}

In \figurename \ref{fig:uallproposedudelayn30k4}, the average MAC delay of the proposed protocol is plotted against normalized offered traffic  for different values of thresholds. As the threshold is increased from $0$ to  $K-1$, the delay decreases as expected. When the threshold is low, the delay is higher because the backoff counter is not allowed to decrement even when the number of ongoing transmissions is small. On the contrary, setting high thresholds increases the probability of collisions, which results in an increased number of dropped packets, thus making an adverse effect on the delay. Thus, the delay under saturated condition for $K_t=3$, is not substantially less than that of the $K_t = 2$ case.

\begin{figure}
\centering
\includegraphics[width=\figwidth]{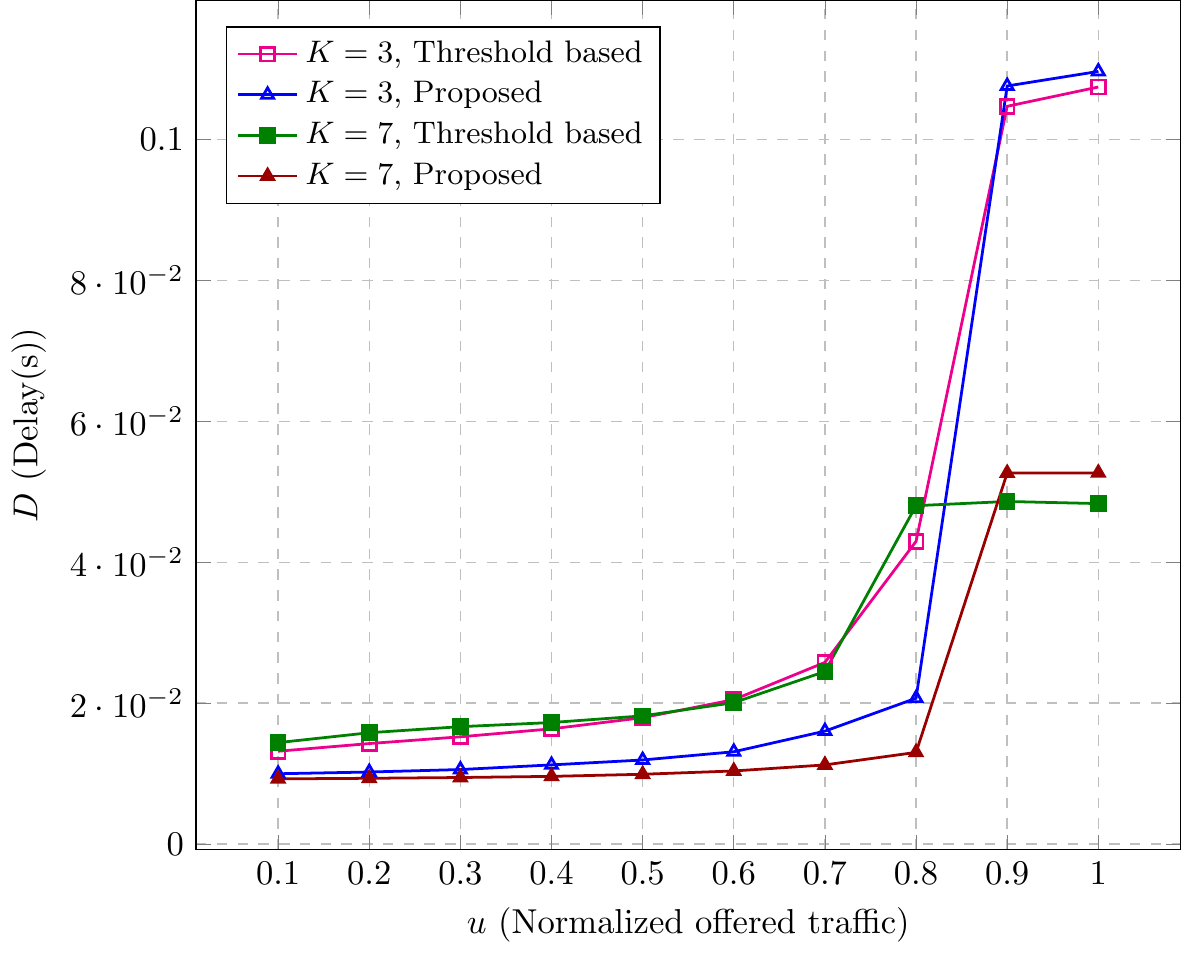}
\caption{The MAC delay of proposed protocol and threshold based protocol against normalized offered traffic (\emph{Params}: Threshold $L_t,K_t=K-1$, Number of stations N = 30, Minimum contention window size $CW_{min}=128$)} 
\label{fig:uall_combinedudelay}
\end{figure}

It can be observed from \figurename \ref{fig:uall_combinedudelay} that the delay of the proposed protocol is less than that of the threshold based protocol.
There is significant improvement in delay for the proposed protocol over threshold based protocol, whenever the total offered traffic is below 80\% of the channel capacity.
When the offered traffic is small, the frequency of encountering idle slots will be more, and the nodes will decrement their backoff counters quickly leading to small delays. Since the backoff counter decrement of the proposed protocol is faster, the delay performance of the proposed protocol is better than that of the threshold based protocol. As the offered traffic increases towards saturation, the difference in the delays between the two protocols reduces. \figurename \ref{fig:uall_combinedudelay} provides a confirmation for our assertion.

\begin{figure}
\centering
\includegraphics[width=\figwidth]{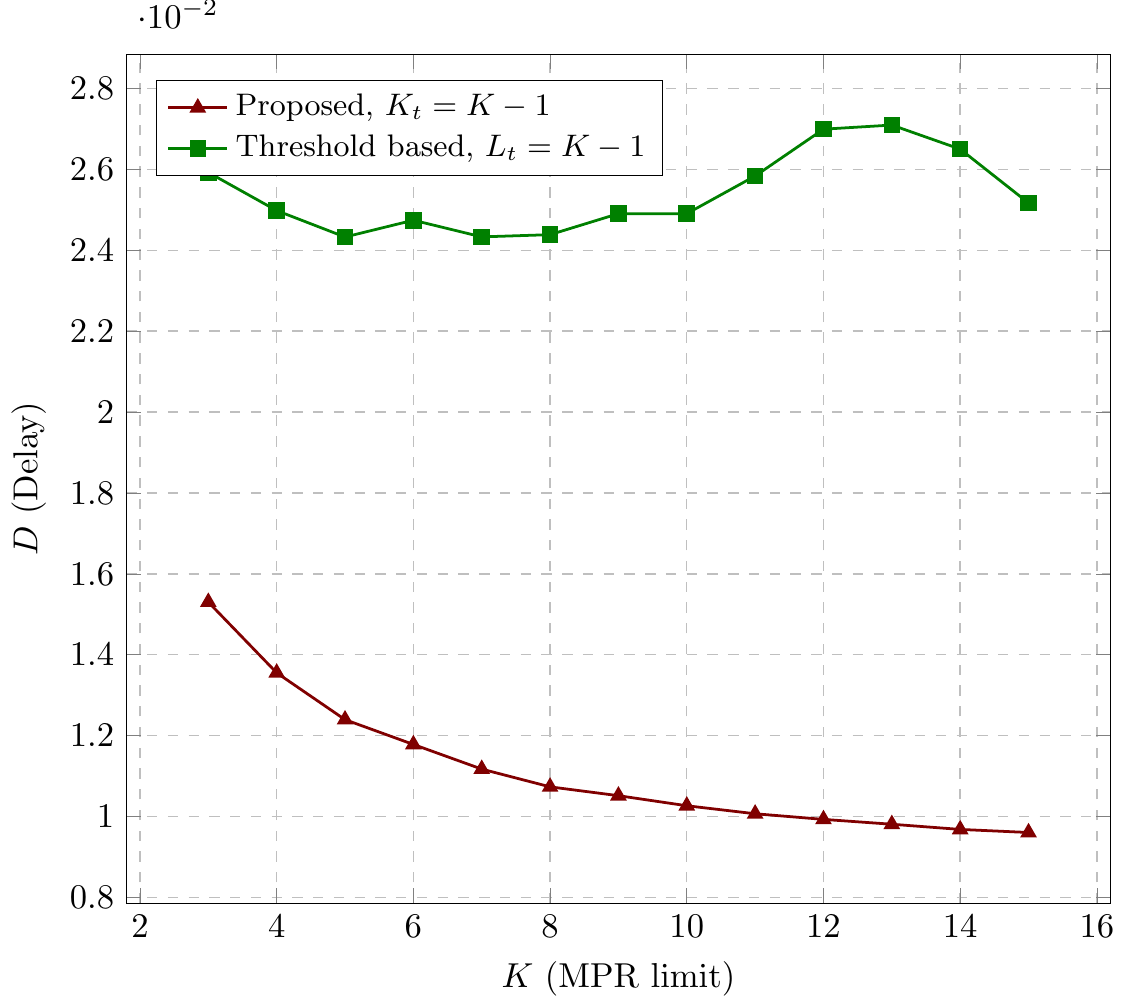}
\caption{The MAC delay of proposed protocol and threshold based protocol against MPR capability $K$ (\emph{Params}: Number of stations $N = 30$, Minimum contention window size $CW_{min}=128$, Normalized offered traffic $u=0.7$)}
\label{fig:uall_combinedkdelayn}
\end{figure}

In \figurename \ref{fig:uall_combinedkdelayn}, variation of MAC delay with MPR capability $K$ is plotted. When the MPR capability is increased in our simulation, we also increase the offered traffic so as to keep the normalized offered traffic constant at the value $0.7$. Clearly, as the MPR capability increases, the delay performance of the proposed protocol scales well.

\subsubsection{Energy Efficiency}

\begin{figure}
\centering
\includegraphics[width=\figwidth]{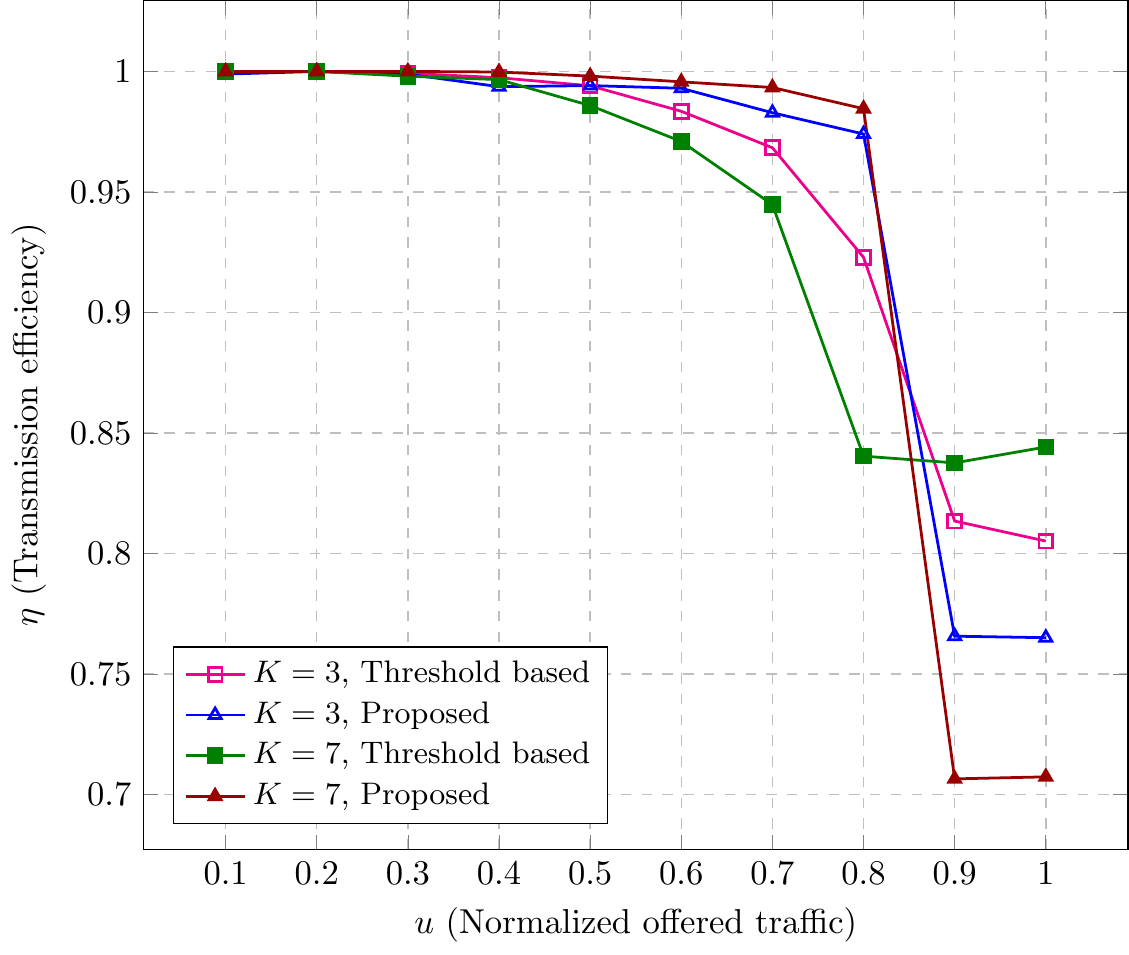}
\caption{The transmission efficiency of proposed protocol and threshold based protocol against normalized offered traffic (\emph{Params}: Thresholds $L_t,K_t=K-1$, Number of stations N = 30, Minimum contention window size $CW_{min}=128$)} 
\label{fig:uall_combinedueff}
\end{figure}

Energy efficiency is very important for sensor network applications. Performance gains at the expense of increased collisions is not desirable for sensor networks.
Since collisions are a major factor for energy consumption, we study the transmissions efficiency as a proportionate metric for studying energy efficiency.
We define transmission efficiency $\eta$ as the inverse of the number of transmission attempts per successful packet transmission. A low value of $\eta$ indicates that the protocol is not energy efficient. In \figurename \ref{fig:uall_combinedueff}, the transmission efficiencies of the two protocols are compared. From the figure, it is clear that the proposed protocol achieves better delay performance without compromising its energy efficiency. \figurename \ref{fig:uall_combinedueff} also shows that under unsaturated traffic conditions, our protocol is more energy efficient than threshold based protocol.

\section{Conclusion}\label{sec:conclusion}
In this paper, we have proposed a simple, backward compatible protocol for MPR wireless channels, which does not require any additional memory or computations. Yet, with this simple design, our protocol achieves significant performance improvement over existing protocols. We have carried out extensive simulations under a wide set of traffic conditions and substantial reduction in MAC delay is obtained for our protocol over IEEE 802.11 DCF and its variant.
The proposed protocol is built around the crucial assumption that the stations are able to accurately determine the number of ongoing transmissions. In reality, the estimate may vary from the actual number of transmissions. The effect of variance of the estimate on the performance of the protocols can be studied. Future work also includes the theoretical performance analysis of the proposed protocol using stochastic models.






%



\bibliographystyle{IEEEtran}
\bibliography{citedin}

\end{document}